\documentclass[11pt]{article} 
\usepackage{graphicx,floatflt,amssymb,epsf,psfig,rotate} 
\def\ltap{\raisebox{-.4ex}{\rlap{$\sim$}} \raisebox{.4ex}{$<$}} 
\def\gtap{\raisebox{-.4ex}{\rlap{$\sim$}} \raisebox{.4ex}{$>$}}

\begin{document} 
\begin{flushright} 
\texttt{hep-ph/0608208}\\ 
SINP/TNP/06-22\\ 
HRI-P-06-08-002 \\
CU-Physics-09/2006
\end{flushright} 
 
\vskip 30pt 
 
\begin{center} 
{\Large \bf Power law scaling in Universal Extra Dimension
scenarios} \\
\vspace*{1cm} 
\renewcommand{\thefootnote}{\fnsymbol{footnote}} 
{ {\sf Gautam Bhattacharyya${}^1$}, {\sf Anindya Datta${}^{2,3}$}, 
{\sf Swarup Kumar Majee${}^{2,3}$}, {\sf Amitava Raychaudhuri${}^{2,3}$} 
} \\ 
\vspace{10pt} 
{\small ${}^{1)}$ {\em Saha Institute of Nuclear Physics, 
1/AF Bidhan Nagar, Kolkata 700064, India} \\
   ${}^{2)}$ {\em Harish-Chandra Research Institute,
Chhatnag Road, Jhunsi, Allahabad  211019, India} \\  
   ${}^{3)}$ {\em Department of Physics, University of Calcutta, 
92 A.P.C. Road, Kolkata 700009, India}}  
 
\normalsize 
\end{center} 
 
\begin{abstract} 
We study the power law running of gauge, Yukawa and quartic scalar couplings
in the universal extra dimension scenario where the extra dimension is
accessed by all the standard model fields. After compactifying on an $S_1
/Z_2$ orbifold, we compute one-loop contributions of the relevant Kaluza-Klein
(KK) towers to the above couplings up to a cutoff scale $\Lambda$. Beyond the
scale of inverse radius, once the KK states are excited, these couplings
exhibit power law dependence on $\Lambda$. As a result of faster running, the
gauge couplings tend to unify at a relatively low scale, and we choose our
cutoff also around that scale. For example, for a radius $R \sim 1\; \rm
TeV^{-1}$, the cutoff is around 30 TeV.  We then examine the consequences of
power law running on the triviality and vacuum stability bounds on the Higgs
mass.  We also comment that the supersymmetric extension of the scenario
requires $R^{-1}$ to be larger than $\sim 10^{10}$ GeV in order that the gauge
couplings remain perturbative up to the scale where they tend to unify.

\vskip 5pt \noindent 
\texttt{PACS Nos:~ 12.60.-i, 11.10.Hi, 11.25.Mj } \\ 
\texttt{Key Words:~~Universal Extra Dimension, Renormalisation group, 
Higgs mass} 
\end{abstract}

\renewcommand{\thesection}{\Roman{section}} 
\setcounter{footnote}{0} 
\renewcommand{\thefootnote}{\arabic{footnote}} 
 
\section{Introduction} 
In the standard model (SM), the gauge, Yukawa and quartic scalar couplings run
logarithmically with the energy scale. Although the gauge couplings do not all
meet at a point, they tend to unify near $10^{15}$ GeV. Such a high scale is
beyond the reach of any present or future experiments. Extra dimensions
accessible to SM fields have the virtue, thanks to the couplings' power law
running, of bringing the unification scale down to an explorable range. Higher
dimensional theories, with radii of compactification around an inverse TeV,
have been investigated from the perspective of high energy experiments,
phenomenology, string theory, cosmology, and astrophysics. Such TeV scale
extra-dimensional scenarios could lead to a new mechanism of supersymmetry
breaking \cite{antoniadis1}, address the issue of fermion mass hierarchy from
a different angle \cite{Arkani-Hamed:1999dc}, provide a cosmologically viable
dark matter candidate \cite{Servant:2002aq}, interpret the Higgs as a quark
composite leading to a successful electroweak symmetry breaking without the
necessity of a fundamental Yukawa interaction \cite{Arkani-Hamed:2000hv}, and,
as mentioned before and what constitutes the central issue of our present
study, lower the unification scale down to a few TeV
\cite{dienes,dienes2}. Our concern here is a specific framework, called the
Universal Extra Dimension (UED) scenario, where there is a single flat extra
dimension, compactified on an $S_1 /Z_2$ orbifold, which is accessed by all
the SM particles \cite{acd}. From a 4-dimensional viewpoint, every field will
then have an infinite tower of Kaluza-Klein (KK) modes, the zero modes being
identified as the SM states.  We examine the cumulative contribution of these
KK states to the renormalisation group (RG) evolution of the gauge, Yukawa and
quartic scalar couplings. Our motive is to extract any subtle features that
emerge due to the KK tower induced power law running of these couplings in
contrast to the usual logarithmic running of the standard 4-dimensional
theories, and whether they set any limit on parameters for the sake of
theoretical and experimental consistency.  Before we illustrate the RG
calculational details, we take a stock of the existing constraints on the UED
scenario, and we comment on what does RG evolution technically mean in the
context of {\em hitherto} non-renormalisable higher dimensional theories.

The key feature of UED is that the momentum in the universal fifth direction
is conserved. From a 4-dimensional perspective this implies KK number
conservation. Strictly speaking, what actually remains conserved is the KK
parity $(-1)^n$, where $n$ is the KK number. As a result, the lightest KK
particle is stable. Also, KK modes cannot affect electroweak processes
at the tree level. They do however contribute to higher order electroweak
processes. In spite of the infinite multiplicity of the KK states, the KK
parity ensures that all electroweak observables are finite (up to
one-loop)\footnote{The observables start showing cutoff sensitivity of various
degree as one goes beyond one-loop or considers more than one extra
dimension.}, and comparison of the observable predictions with experimental
data yields bounds on $R$.  Constraints on the UED scenario from $g-2$ of the
muon \cite{nath}, flavour changing neutral currents \cite{chk,buras,desh}, $Z
\to b\bar{b}$ decay \cite{santa}, the $\rho$ parameter \cite{acd,appel-yee},
several other electroweak precision tests \cite{ewued} and implications from
hadron collider studies \cite{collued}, all conclude that $R^{-1}~\gtap~300$
GeV.

We now come to the technical meaning of RG running in a higher
dimensional context. This has been extensively clarified in
\cite{dienes} in a general context, and here we merely reiterate it to
put our specific calculations into perspective. Like all other
extra-dimensional models, from a 4-dimensional point of view, the UED
scenario too is non-renormalisable due to the infinite multiplicity of
the KK states\footnote{For a study of ultraviolet cutoff sensitivity
in different kinds of TeV scale extra-dimensional models, see
\cite{db}.}. So `running' of couplings as a function of the energy scale $\mu$
ceases to make sense. What we should say is that the couplings receive finite
quantum corrections whose size depend on some explicit
cutoff\footnote{The beta functions are coefficients of the
divergence $1/\epsilon$ in a 4-dimensional theory. Here, a second
kind of divergence appears when the finite beta functions get
corrections from each layer of KK states which are summed over.
This summation is truncated at a scale $\Lambda$.} $\Lambda$. The
corrections originate from the $\Lambda R$ number of KK states
which lie between the scale $R^{-1}$ where the first KK states
are excited and the cutoff scale $\Lambda$. The couplings will
have a power law dependence on $\Lambda$ as a result of the KK
summation.  This cutoff is interpreted as the scale where a
paradigm shift occurs when some new renormalisable physics
underlying our effective non-renormalisable framework surfaces.

\section{Universal Extra Dimension}
The extra dimension is compactified on a circle of radius $R$ with a $Z_2$
orbifolding identifying $y \to -y$, where $y$ denotes the fifth compactified
coordinate. The orbifolding is crucial in generating
{\em chiral} zero modes for fermions. After integrating out the compactified
dimension, the 4-dimensional Lagrangian can be written involving the zero mode
and the KK modes.  To appreciate the contributions of the KK towers into the
so-called RG evolutions, it is instructive to have a glance at the KK mode
expansions of these fields. Each component of a 5-dimensional field must be
either even or odd under the orbifold projection. The KK expansions are given
by,
\begin{eqnarray}
\label{fourier}
A_{\mu}(x,y)&=&\frac{\sqrt{2}}{\sqrt{2\pi
R}}A_{\mu}^{(0)}(x)+\frac{2}{\sqrt{2\pi
R}}\sum^{\infty}_{n=1}A_{\mu}^{(n)}(x)\cos\frac{ny}{R},~~~~
A_5(x,y) = \frac{2}{\sqrt{2\pi
R}}\sum^{\infty}_{n=1}A_5^{(n)}(x)\sin\frac{ny}{R}, \nonumber\\
\phi(x,y)&=&\frac{\sqrt{2}}{\sqrt{2\pi
R}}\phi^{(0)}(x)+\frac{2}{\sqrt{2\pi
R}}\sum^{\infty}_{n=1}\phi^{(n)}(x)\cos\frac{ny}{R}, \nonumber \\
\mathcal{Q}_{i}(x,y)&=&\frac{\sqrt{2}}{\sqrt{2\pi 
R}}\bigg[{\pmatrix{u_i\cr d_i}}_{L}(x)+\sqrt{2}\sum^{\infty}_{n=1}\Big[
\mathcal{Q}^{(n)}_{iL}(x)\cos\frac{ny}{R}+
\mathcal{Q}^{(n)}_{iR}(x)\sin\frac{ny}{R}\Big]\bigg], \\
\mathcal{U}_{i}(x,y)&=&\frac{\sqrt{2}}{\sqrt{2\pi
R}}\bigg[u_{iR}(x)+\sqrt{2}\sum^{\infty}_{n=1}\Big[
\mathcal{U}^{(n)}_{iR}(x)\cos\frac{ny}{R}+
\mathcal{U}^{(n)}_{iL}(x)\sin\frac{ny}{R}\Big]\bigg], \nonumber\\
\mathcal{D}_{i}(x,y)&=&\frac{\sqrt{2}}{\sqrt{2\pi
R}}\bigg[d_{iR}(x)+\sqrt{2}\sum^{\infty}_{n=1}\Big[
\mathcal{D}^{(n)}_{iR}(x)\cos\frac{ny}{R}+
\mathcal{D}^{(n)}_{iL}(x)\sin\frac{ny}{R}\Big]\bigg], \nonumber
\end{eqnarray}
where $i=1,2,3$ are generation indices. Above, $x (\equiv x^{\mu})$ denotes
the first four coordinates, and as mentioned before, $y$ is the compactified
coordinate. The complex scalar field $\phi (x,y)$ and the gauge boson
$A_\mu(x,y)$ are $Z_2$ even fields with their zero modes identified with the
SM scalar doublet and a SM gauge boson respectively. On the contrary, the
field $A_5(x,y)$, which is a real scalar transforming in the adjoint
representation of the gauge group, does not have any zero mode. The fields
$\mathcal{Q}$, $\mathcal{U}$, and $\mathcal{D}$ describe the 5-dimensional
quark doublet and singlet states, respectively, whose zero modes are
identified with the 4-dimensional chiral SM quark states. The KK expansions of
the weak-doublet and -singlet leptons will be likewise and are not shown for
brevity.

\section{Renormalisation Group Equations} 
We now lay out the strategy followed to compute the RG correction
to the couplings from the KK modes. The first step is obviously
the calculation of the contribution from a given KK level which
has both $Z_2$-even and -odd states. Three points are noteworthy
and should be taken into consideration during this step:
\begin{enumerate} 
\item While the zero mode fermions are chiral as a result of
orbifolding, the KK quarks and leptons at a given level are
vector-like. 
\item The fifth compotent of the gauge bosons are ($Z_2$ odd)
scalars\footnote{The coupling of the $A_5^{(n)}$ states to
fermions involve $\gamma_5$ and so, strictly, they are pseudoscalars.}, but in
the adjoint representation of the gauge group. Such states are not encountered
in the SM context.
\item The KK index $n$ is conserved at each tree level vertex.
\end{enumerate}
The first step KK excitation occurs at the scale $R^{-1}$ (modulo the zero
mode mass). Up to this scale the RG evolution is logarithmic, controlled by
the SM beta functions. Between $R^{-1}$ and $2 R^{-1}$, the running is still
logarithmic but with beta functions modified due to the first KK level
excitations, and so on. Every time a KK threshold is crossed, new resonances
are sparked into life, and new sets of beta functions rule till the next
threshold arrives. The beta function contributions are the same for each of
the $\Lambda R$ KK levels, which, in effect, can be summed. After this, the
scale dependence is not logarithmic any more, it shows power law behaviour, as
illustrated by Dienes {\em et al} in \cite{dienes2}. This illustration shows
that if $\Lambda R \gg 1$, then to a very good accuracy the calculation
basically boils down to computing the number of KK states up to the cutoff
scale. For one extra dimension up to the energy scale $E$ this number is $S =
ER$, and $E^{\rm max} = \Lambda$. Then if $\beta^{\rm SM}$ is a generic SM
beta function valid during the logarithmic running up to $R^{-1}$, beyond that
scale one should replace it as\footnote{We refer the readers to Eqs.~(2.15)
and (2.21) of Ref.~\cite{dienes2}, and the subtleties leading to these
equations in the context of gauge couplings, to have a feel for our
Eq.~(\ref{master}).}
\begin{equation} 
\label{master}
\beta^{\rm SM} \to \beta^{\rm SM} + (S - 1) \tilde{\beta}, 
\end{equation}
where $\tilde{\beta}$ is a generic contribution from a single KK
level. Irrespective of whether we deal with the `running' of gauge, Yukawa, or
quartic scalar couplings, the structure of Eq.~(\ref{master}) would continue
to hold. Clearly, the $S$ dependence reflects power law running. How this
master formula (\ref{master}) enters diagram by diagram into the evolution of
the above couplings in the UED scenario constitutes the main part of
calculation in the present paper.

\subsection{Gauge couplings} 
While considering the evolution of the gauge couplings, we first write
$\tilde{\beta}_i^g = \tilde{b}_i g_i^3$. The calculation of $\tilde{b}_i^g$
would proceed via the same set of Feynman graphs which give the SM
contributions $b_i^{\rm SM}$ but now containing the KK internal lines.  The
key points to remember are the presence of adjoint scalars and doubling of KK
quark and lepton states due to their vectorial nature.

%%%%%%%%%%%%%%%%%%%%%%%%%%%%%%%%%%%%%%%%%%%%%%%%%%%%%%%%%%%%%%%%%%%%%%%%%%%
  
\begin{center}
\begin{figure}[thb]
\hspace*{2cm}\psfig{figure=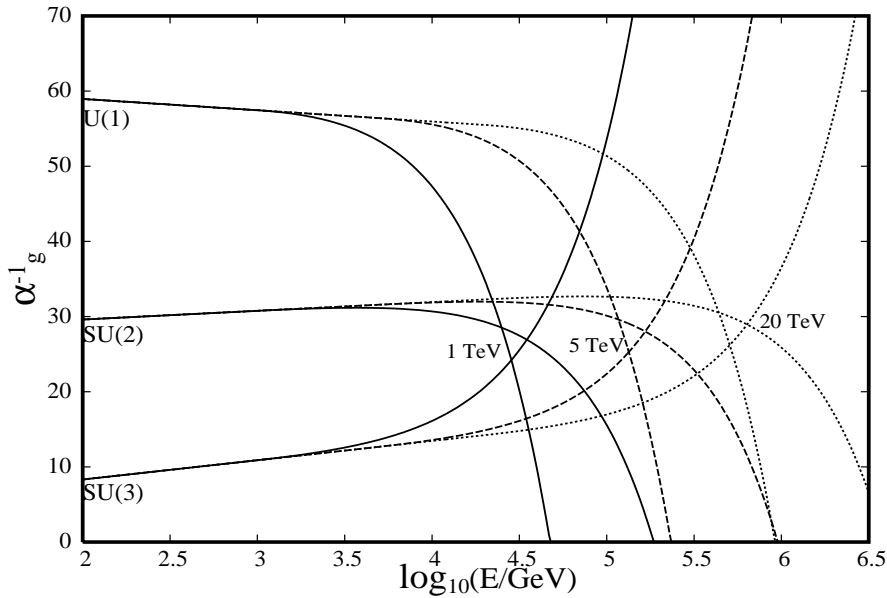,width=12cm,height=8cm,angle=270}
\caption{\sf \small {Evolution of gauge couplings for UED with $R^{-1}$ = 1,
5, and 20 TeV. For each of the three couplings, $\alpha_g \equiv g^2/4\pi$.}}
\label{f:gunif}
\end{figure}
\end{center}
%%%%%%%%%%%%%%%%%%%%%%%%%%%%%%%%%%%%%%%%%%%%%%%%%%%%%%%%%%%%%%%%%%%%%%%%%%%

We obtain
\begin{equation}
\tilde{b}_1 = \frac{81}{10}, ~~ \tilde{b}_2 = \frac{7}{6}, ~~ 
\tilde{b}_3 = - \frac{5}{2},
\label{ued}
\end{equation}   
where the U(1) beta function is appropriately normalised. Just to
recall, the corresponding SM numbers are 41/10, $-$19/6, $-$7,
respectively. We have plotted the evolution of gauge couplings in UED
for $R^{-1} =$ 1, 5, and 20 TeV in Fig.~\ref{f:gunif}. The running is
fast, as expected, and the couplings nearly meet around\footnote{The
issue of proton stability in such low scale unification scenarios has
been dealt in \cite{pdecay}.}  30, 138 and 525 TeV,
respectively.  It is not hard to provide an intuitive argument for
such low unification scales and how they vary with $R$: roughly
speaking, $\Lambda R$ is order $\ln (M_{\rm GUT}/M_W) \sim
\ln(10^{15})$, where $M_{\rm GUT}$ is the 4-dimensional GUT scale,
i.e. the effect of a slow logarithmic running over a large scale is
roughly reproduced by a fast power law sprint over a short track.  The
other striking feature reflected in Fig.~\ref{f:gunif} is that the
SU(2) gauge coupling ceases to be asymptotically free: the dominance
of the KK matter sector over the gauge part in $\tilde{b}_2$ severely
challenges the SU(2) asymptotic freedom.  In contrast, the negative
sign of $\tilde{b}_3$ causes a precipitous drop in the SU(3) gauge
coupling with energy.

\subsection{Yukawa couplings} 
%%%%%%%%%%%%%%%%%%%%%%%%%%%%%%%%%%%%%%%%%%%%%%%%%%%%%%%%%%%%%%%%%%
\begin{center}
\begin{figure}[thb]
\hspace*{2cm}\psfig{figure=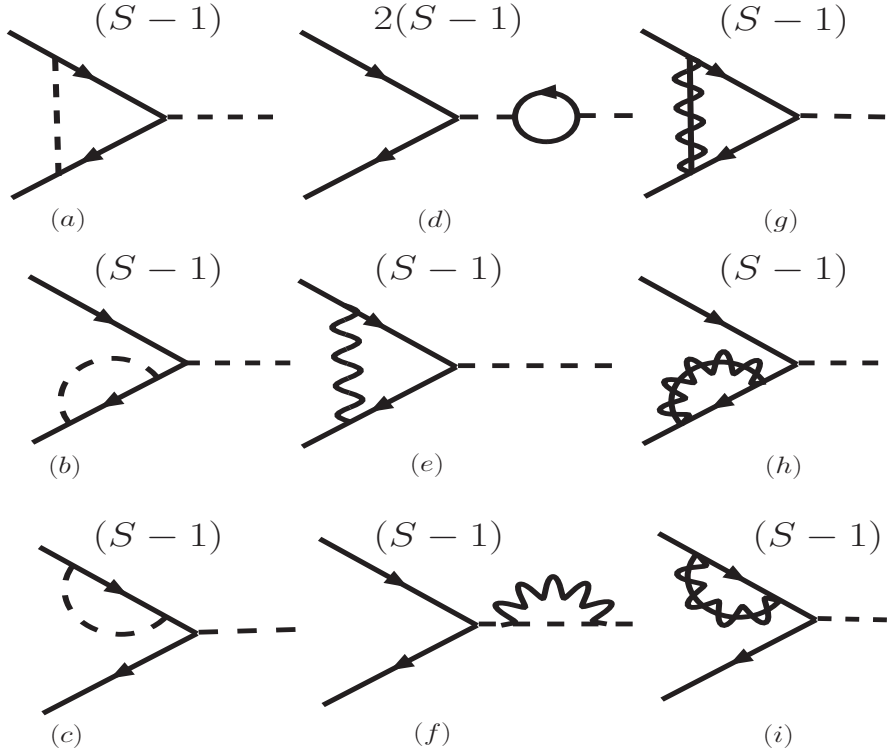,width=12cm,height=10cm,angle=0}
\caption{\sf \small {Diagrams contributing to Yukawa coupling
evolution in the Landau gauge. Solid (broken) lines correspond to
fermions (SM scalars), while wavy lines (wavy+solid lines) represent
ordinary gauge bosons (fifth components of gauge bosons).}}
\label{f:yuk}
\end{figure}
\end{center}
%%%%%%%%%%%%%%%%%%%%%%%%%%%%%%%%%%%%%%%%%%%%%%%%%%%%%%%%%%%%%%%%%%%%
%%%%%%%%%%%%%%%%%%%%%%%%%%%%%%%%%%%%%%%%%%%%%%%%%%%%%%%%%%%%%%%%%%%
\begin{center}
\begin{figure}[thb]
%\hspace*{2cm}\psfig{figure=yukawa.eps,width=12cm,height=8cm,angle=0}
%\vskip -30pt
\hskip 1.50cm
\psfig{figure=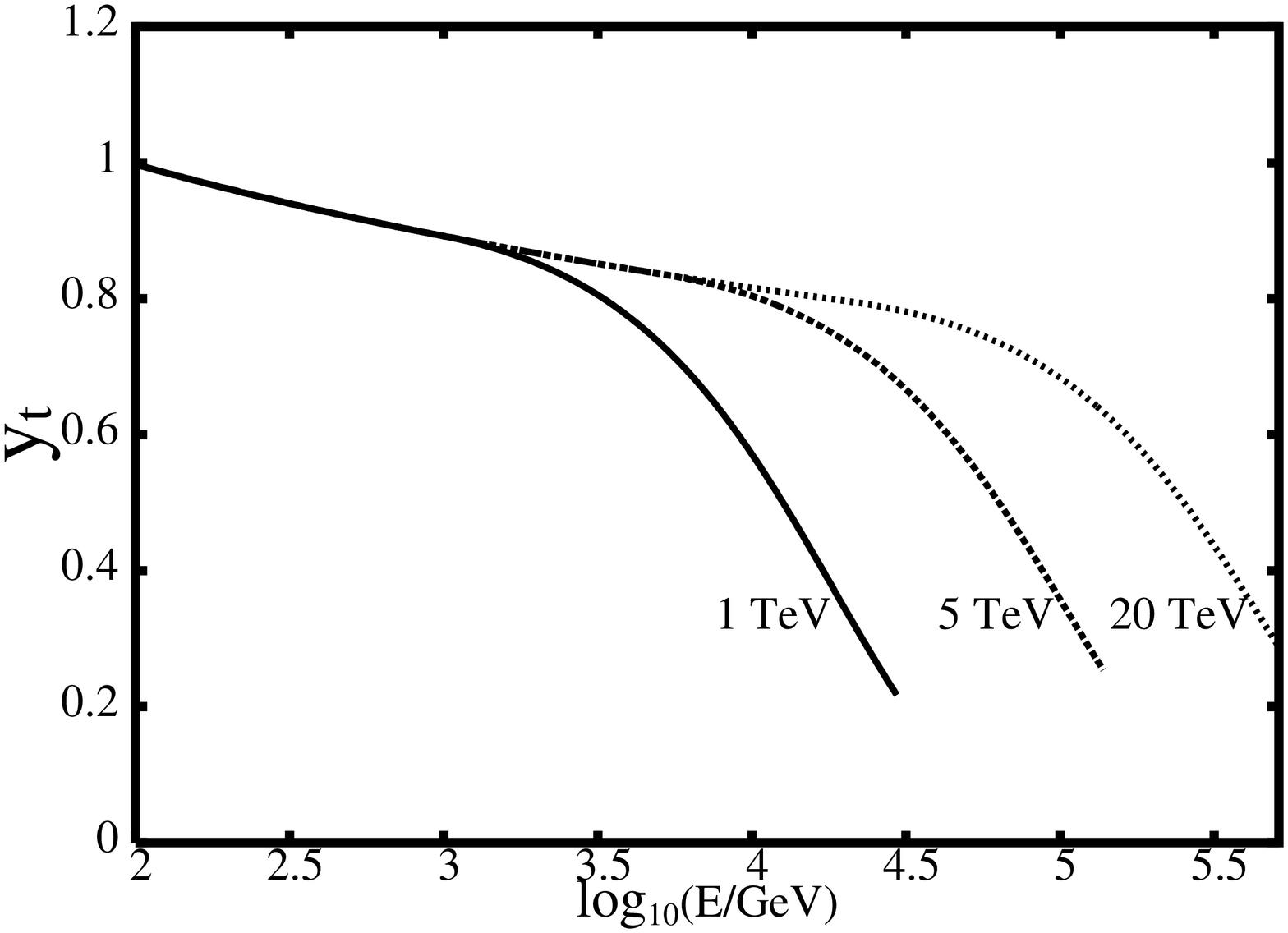,width=6.2cm,height=6.2cm,angle=270}
\vskip -6.23cm
\hskip 8.0cm
\psfig{figure=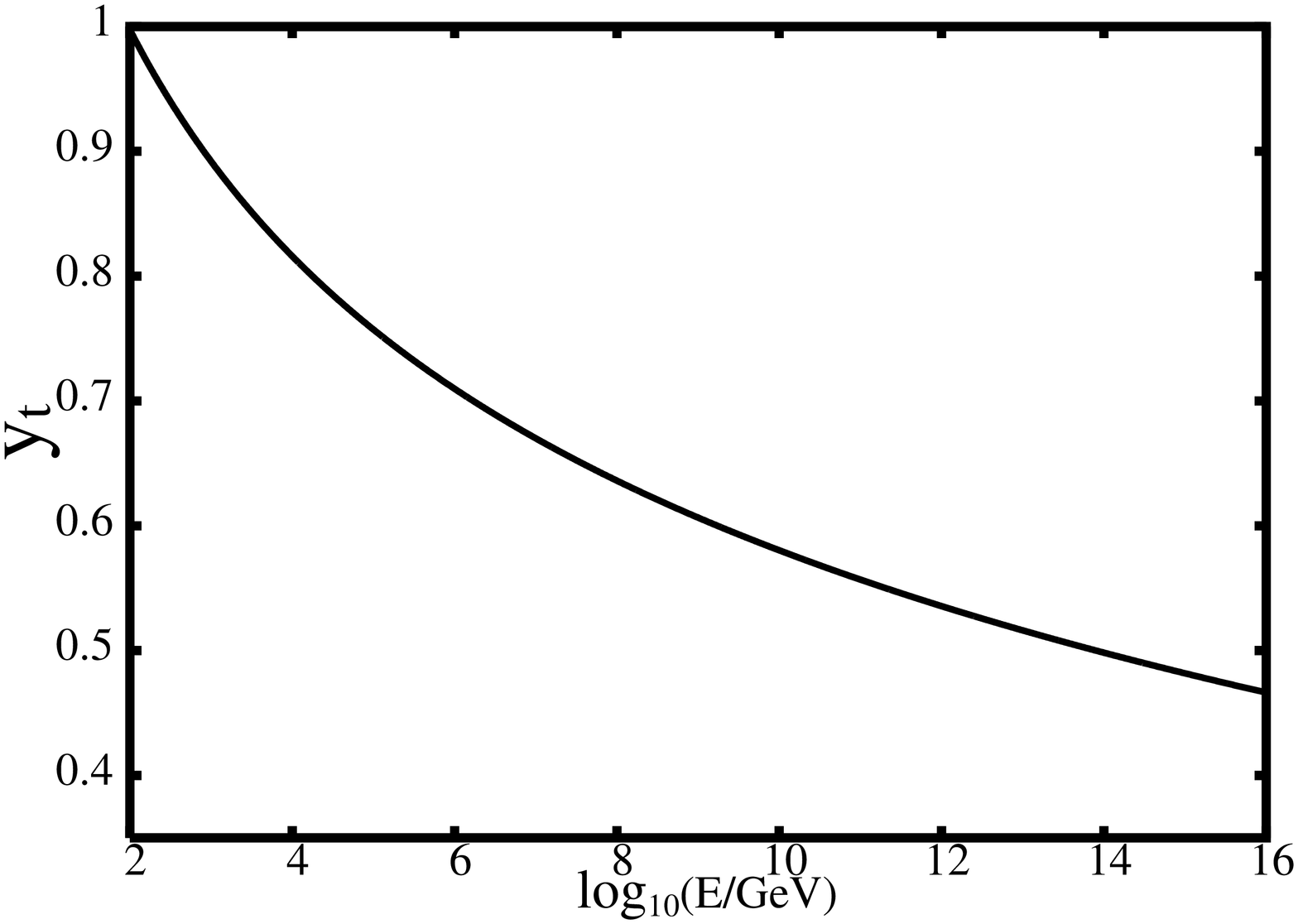,width=6.2cm,height=6.2cm,angle=270}
\caption{\sf \small {Evolution of the top quark Yukawa coupling in the UED
scenario (left panel) and (b) the SM (right panel). UED evolution is shown for
three different values of $R^{-1}$ and the curves are terminated at the
corresponding unification scales.}}
\label{f:tyuk}
\end{figure}
\end{center}
%%%%%%%%%%%%%%%%%%%%%%%%%%%%%%%%%%%%%%%%%%%%%%%%%%%%%%%%%%%%%%%%%%%
The Feynman diagrams that contribute to the power law evolution of
Yukawa couplings (in Landau gauge) are shown in Fig.~\ref{f:yuk}.  The
contributions come from the pure SM states, their KK towers, and from
the adjoint representation scalars\footnote{A subtle feature is worth
noticing. In four dimensions, the calculational advantage of working
in Landau gauge is that some diagrams give vanishing
contributions. The argument breaks down in a higher dimensional
context. More explicitly, consider the Figs.~\ref{f:yuk}h and
\ref{f:yuk}i. These graphs proceed through the exchange of adjoint
$A_5$ scalars and yield non-vanishing contributions. The corresponding
figures with $A_\mu$ exchange are absent because they give null
results in the Landau gauge.}.  The last two contributions, as the
master formula (\ref{master}) indicates, have an overall
proportionality factor $(S-1)$. As we examine contributions from
individual KK states, we see that due to the argument of fermion
chirality, not in all diagrams do the cosine and sine mode states both
{\em simultaneously} contribute.  This accounts for a relative factor
of 2 between the two types of diagrams.  For example, in
Fig.~\ref{f:yuk}a the fermionic KK modes can only come from cosine
expansions, whereas in Fig.~\ref{f:yuk}d both cosine and sine fermion
modes contribute. This is why Fig.~\ref{f:yuk}a has a multiplicating
factor $(S-1)$, while for Fig.~\ref{f:yuk}d the factor is $2(S-1)$.
Whereever $A_5$ is involved as an internal line, the associated KK
internal fermions necessarily come from sine expansion, e.g. in
Figs.~2g, 2h and 2i.  The above book-keeping has been done for
individual graphs and the proportionality factors have been mentioned
for each diagram in Fig.~\ref{f:yuk}. The Yukawa RG equations (beyond
the threshold $R^{-1}$) can be written as ($t = \ln E$):
\begin{equation}
16 \pi^2 \frac{dy_f}{dt} = \beta^{\rm SM}_{y_f} +   \beta^{\rm UED}_{y_f},
\label{eq:gen}
\end{equation}
where $f$ generically stands for the up/down quarks or leptons. The SM
beta functions $\beta^{\rm SM}_{y_f}$ can be found e.g. in
\cite{sumathi}.  The UED contributions to the beta functions
$\beta^{\rm UED}_{y_{l,u,d}}$ are given by:
\vbox{
\begin{eqnarray}
\beta^{\rm UED}_{y_{l}} &=& (S -
1)\;\left[-(\frac{21}{8}g_2^2 +\frac{129}{40}g_1^2 )
  +\frac{3}{2} y_l^2 \right]\;y_l + 2(S -
1)\;\left[Y_l + 3 Y_u + 3 Y_d \right]\;y_l ,\nonumber \\
\beta^{\rm UED}_{y_{u}} &=& (S - 1)\;\left[-(12 g_3^2 +
  \frac{21}{8}g_2^2 +\frac{9}{8}g_1^2 ) + \frac{3}{2}(y_u^2 -
  y_d^2)\right]\;y_u + 2(S -
1)\;\left[Y_l + 3 Y_u + 3 Y_d \right]\;y_u ,   \\
\beta^{\rm UED}_{y_{d}} &=& (S - 1)\;\left[-(12 g_3^2 +
  \frac{21}{8}g_2^2 + \frac{9}{40}g_1^2 ) + \frac{3}{2}(y_d^2 -
  y_u^2)\right]\;y_d + 2(S - 1)\;\left[Y_l + 3 Y_u + 3 Y_d \right]\;
y_d , \nonumber
\end{eqnarray}
}
with $Y_l = \sum_{l} y_l^2$, $Y_d = \sum_{d} y_d^2$, and $Y_u =
\sum_{u} y_u^2$. 
To illustrate how the power law dependence of Yukawa couplings quantitatively
compares and contrasts with their 4-dimensional logarithmic running, we have
exhibited in Fig.~\ref{f:tyuk} the behaviour of the top-quark Yukawa
coupling in the two cases. 

Another consequence of unification in many models is a prediction
of the low energy value of $m_b/m_\tau$. This ratio, unity at
the unification scale, at low energies takes the values 4.7, 4.2,
and 3.9 for $1/R$ = 1, 5, and 20 TeV, respectively. Admittedly,  $m_b$ is 
on the high side; a limitation which perhaps may be attributable
to the one-loop level of the calculation.

%%%%%%%%%%%%%%%%%%%%%%%%%%%%%%%%%%%%%%%%%%%%%%%%%%%%%%%%%%%%%%%%%%%%%
\begin{center}
\begin{figure}[thb]
\hspace*{3cm}\psfig{figure=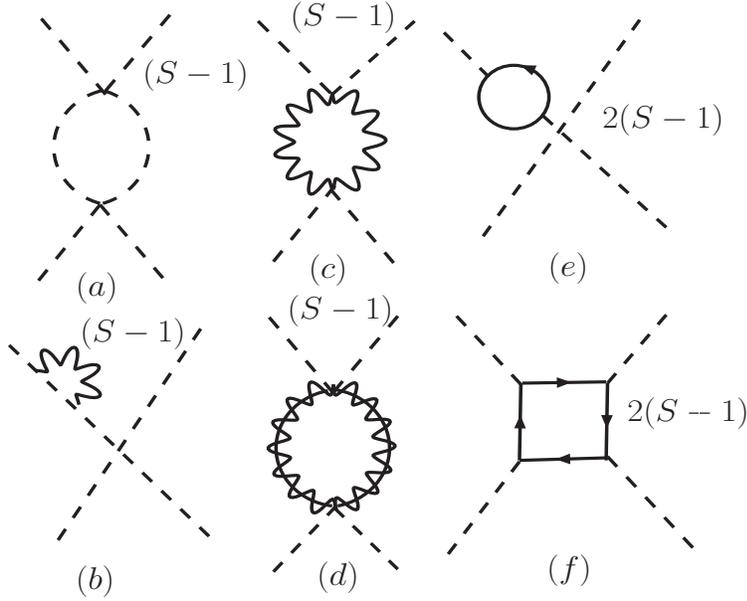,width=10cm,height=8cm,angle=0}
\caption{\sf \small {Diagrams contributing to quartic scalar
coupling evolution. The conventions are the same as in Fig. \ref{f:yuk}.}}
\label{f:lamb}
\end{figure}
\end{center}
%%%%%%%%%%%%%%%%%%%%%%%%%%%%%%%%%%%%%%%%%%%%%%%%%%%%%%%%%%%%%%%%%%%%%%
%%%%%%%%%%%%%%%%%%%%%%%%%%%%%%%%%%%%%%%%%%%%%%%%%%%%%%%%%%%%%%%%%%%%%%%
\begin{center}
\begin{figure}[thb]
\hspace*{2cm}\psfig{figure=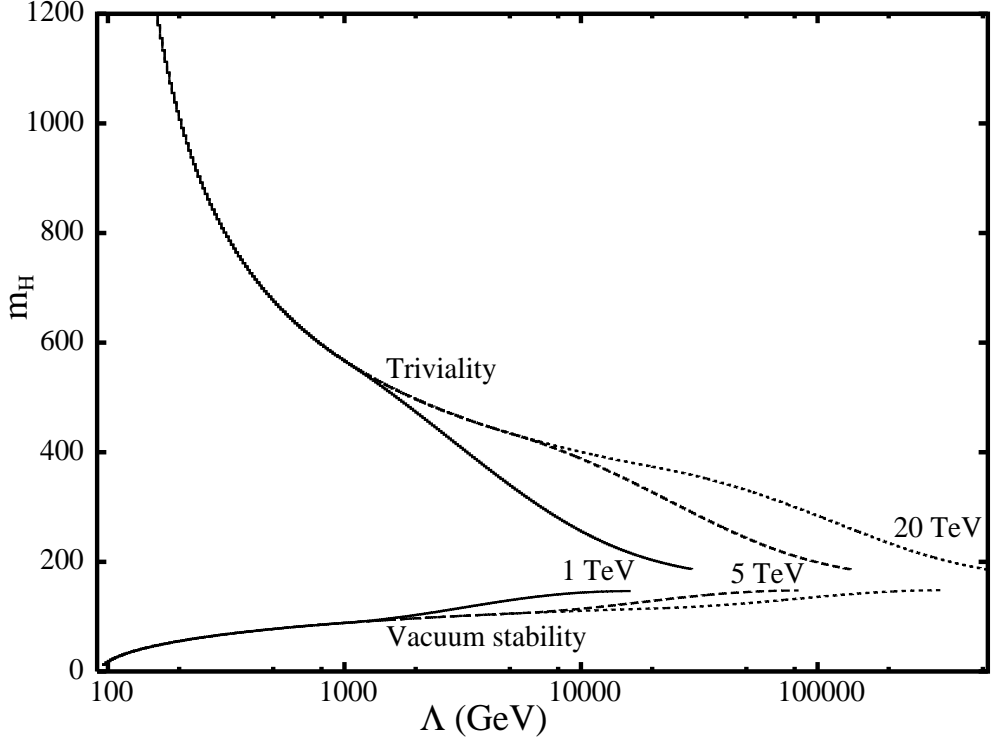,width=14cm,height=10cm,angle=270}
\caption{\sf \small {Bounds on the Higgs mass at the electroweak scale in the
UED scenario from the triviality and vacuum stability conditions for $R^{-1}$
= 1, 5, and 20 TeV. See text for details.}}
\label{f:higgs}
\end{figure}
\end{center}
%%%%%%%%%%%%%%%%%%%%%%%%%%%%%%%%%%%%%%%%%%%%%%%%%%%%%%%%%%%%%%%%%%%%%%%
\subsection{Quartic scalar coupling and the Higgs mass}
The one-loop diagrams through which the KK modes contribute to the power law
running of the quartic scalar coupling $\lambda$ (in Landau gauge) are shown
in Fig.~\ref{f:lamb}.  As clarified before in the case of Yukawa running, the
extra factor of 2 in front of $(S - 1)$ for some graphs indicates that
cosine and sine KK modes {\em both} contribute only to those graphs. The
evolution equation can be written as
\begin{equation} 
16 \pi^2 \frac{d\lambda}{dt} = \beta^{\rm SM}_{\lambda} +
\beta^{\rm UED}_{\lambda} 
\end{equation}

The expressions for $\beta^{\rm SM}_{\lambda}$ can be found e.g.~in
\cite{chengli}.  The UED beta functions are given by
\vbox{
\begin{eqnarray}
\beta^{\rm UED}_{\lambda} &=&
(S - 1)\left[ 3 g_2^4 + \frac{6}{5}g_2^2 g_1^2 + \frac{9}{25}
g_1^4 -3 \lambda (3 g_2^2 +
\frac{3}{5}g_1^2) + 12 \lambda^2 \right] \nonumber \\
&+& 2(S - 1)\left[  4 \left(Y_l + 3 Y_u +
    3 Y_d \right) \lambda -4\sum_{l,u,d}\left(y_l^4 + 3 y_u^4 + 3 y_d^4
  \right) \right].  
\end{eqnarray}
} The evolution of $\lambda$ has interesting bearings on the Higgs mass. In
the standard 4-dimensional context, bounds on the Higgs mass have been placed
on the grounds of `triviality' and `vacuum stability' \cite{trivia}. What do
they imply in the UED context? The `triviality' argument requires that
$\lambda$ stays away from the Landau pole, i.e. remains finite, all the way to
the cutoff scale $\Lambda$. The condition that $1/\lambda(\Lambda) > 0$ can be
translated to an upper bound on the Higgs mass ($m_H$) at the electroweak
scale when the cutoff of the theory is $\Lambda$. This has been plotted in
Fig.~\ref{f:higgs} (the upper curves) for three different values of $R$.  A
given point on that curve (for a given $R$) corresponds to a maximum allowed
$m_H$ at the weak scale; for a larger $m_H$ the coupling $\lambda$ becomes
infinite at some scale less than $\Lambda$ and the theory ceases to be
perturbative.  Clearly, this $m_H^{\rm max}$ varies as we vary the cutoff
$\Lambda$ . The argument of `vacuum stability' relies on the requirement that
the scalar potential be always bounded from below, i.e. $\lambda (\Lambda) >
0$. This can be translated to a lower bound $m_H^{\rm min}$ at the weak
scale. The lower set of curves in Fig.~\ref{f:higgs} (for three values of
$R^{-1}$) represent the `vacuum stability' limits, the region below the curve
for a given $R$ being ruled out. Recalling that the cutoff is where the gauge
couplings tend to unify, we observe that the Higgs mass is limited in the
narrow zone
\begin{equation}
148 ~ \ltap~ m_H ~\ltap~ 186 ~{\rm GeV}
\label{higgsued} 
\end{equation}
in all the three cases, for a zero mode top quark mass of 174.2
GeV. Admittedly, our limits are based on one-loop corrections
only. That the upper and lower limits are insensitive to the choice of
$R$ is not difficult to understand, as what really counts is the
number of KK states, given by the product $\Lambda R$, which, as
mentioned before, is nearly constant, order $\ln (10^{15})$.  The
limits in Eq.~(\ref{higgsued}) are very close to what we obtain in the
SM at the one-loop level, namely $147 ~\ltap ~ m_H^{\rm SM} ~
\ltap 189$ GeV (see also \cite{Kielanowski:2003jg}, where one-loop SM results
have been derived\footnote{The SM two-loop limits are \cite{trivia}: $145
~\ltap ~ m_H^{\rm SM} ~ \ltap~ 168$ GeV for $m_t = 174.2$ GeV.}).

\subsection{Supersymmetric UED}  
What happens if we take the supersymmetric (SUSY) version of UED?  A
5-dimensional $N=1$ supersymmetry when perceived from a 4-dimensional context
contains two different $N=1$ multiplets forming one $N=2$ supermultiplet.  For
a comprehensive analysis, we refer the readers to \cite{dienes}.  There are
two issues that immediately concern our analysis. First, unlike in the
non-SUSY case, the Higgs scalar in a chiral multiplet will now have both even
and odd $Z_2$ modes on account of degrees of freedom counting consistent with
supersymmetry. Also, there will be two such $N=2$ chiral supermultiplets to
meet the requirement of supersymmetry. Second, in the RG evolution two energy
scales will come into play. The first of these is the supersymmetry scale,
called $M_S$, which we take to be 1 TeV. Beyond $M_S$, supersymmetric
particles get excited and their contributions must be included in the RG
evolution. The second scale is that of the compactified extra dimension $1/R$,
which we take to be larger than $M_S$.

The gauge coupling evolution must now be specified for three
different regions. The first of these is when $E < M_S$ where the
SM with the additional scalar doublet\footnote{SUSY requires two
complex scalar doublets.}  beta functions are in control. In this region:
\begin{equation}
{b}_{1o} = \frac{21}{5}, ~~ {b}_{2o} = -\frac{10}{3}, ~~ 
{b}_{3o} = -7 .
\label{smplus}
\end{equation}
Once $M_S$ is crossed and up until $1/R$, we also have the superpartners of
the SM particles pitching in with their effects. The contributions of the SM
particles and their superpartners together are given by:
\begin{equation}
{b}_{1s} = \frac{33}{5}, ~~ {b}_{2s} = 1, ~~ 
{b}_{3s} = -3 .
\label{susy1}
\end{equation}
Finally, when the KK-modes are excited ($E > 1/R$) one has 
further contributions from the individual modes: 
\begin{equation}
\tilde{b}_1 = \frac{66}{5}, ~~ \tilde{b}_2 = 10, ~~ 
\tilde{b}_3 = 6 .
\label{susy2}
\end{equation}   
Thus, beyond $1/R$, the total contribution is given by 
\begin{equation} 
b_i^{\rm tot} = b_{io}  + \Theta (E-M_S) ~(b_{is} - b_{i0}) + \Theta (E-
\frac{1}{R}) ~(S - 1)  ~\tilde{b}_i, 
\end{equation}

Not unexpectedly, for the SUSY UED case, gauge unification is possible.  We
observe that the introduction of this plethora of KK excitations of the SM
particles and their superpartners radically changes the beta functions; so
much so, that the gauge couplings tend to become non-perturbative before
unification is achieved. For clarity, we make the argument more explicit
below. First, from Eqs.~(\ref{susy1}) and (\ref{susy2}) we note that the
dominance of the KK matter over the KK gauge parts is so overwhelming that the
SU(3) beta function ($\tilde{b}_3$) beyond the first KK threshold ceases to be
negative any longer. The other two gauge beta functions, which were already
positive with contributions from zero mode particles plus their superpartners,
become even more positive. So the curves for all the three gauge couplings
would have the same sign slopes once the KK modes are excited. As a result,
with increasing energy the three curves for $\alpha_g^{-1}$ would dip with a
power law scaling fast into a region where the couplings themselves become too
large at the time they meet.  Therefore, in order that all of them remain
perturbative during the entire RG evolution, the onset of the KK dynamics has
to be sufficiently delayed. This requirement imposes $R^{-1}~\gtap~5.0 \times
10^{10}$ GeV.  In effect, this implies that the twin requirements of a
SUSY-UED framework as well as perturbative gauge coupling unification pushes
the detectability of the KK excitations well beyond the realm of the LHC.

\section{Conclusions and Outlook}
As the LHC is getting all set to roar in 2007, expectations are
mounting as we prepare ourselves to get a glimpse of new and
unexplored territory. New physics of different incarnations,
especially supersymmetry and/or extra dimensions, are crying out for
verification. How does the landscape beyond the electroweak scale
confront the evolution of the gauge, Yukawa and scalar quartic
couplings? Will there be a long logarithmic march through the desert
all the way to $10^{(16-17)}$ GeV, or is a power law sprint 
awaiting us with a stamp of extra dimensions? In which way does the
latter quantitatively differ from the former has been the subject of
our investigation in the present paper. We observe the following
landmarks that characterise the extra-dimensional running:
\begin{enumerate}
\item
The orbifolding renders some subtle features to the RG running in UED.  Due to
the conservation of KK number at tree level vertices, the $Z_2$ even and odd
KK states selectively contribute to different diagrams. While some diagrams
are forbidden, there are new diagrams originating from adjoint scalar
exchanges. In the present article we have performed a diagram by
diagram book-keeping leading to the evolution equations.

\item Low gauge coupling unification scales can be achieved without
introducing non-perturbative gauge couplings. The unification scale depends on
$R$, and is approximately given by $\Lambda \sim (25-30)/R$.

\item The `triviality' and `vacuum stability' bounds on the Higgs mass have
  been studied in the context of power law evolution. This limits the
  Higgs mass in the range $148 ~ \ltap m_H ~\ltap 186$ GeV at the
  one-loop level. The corresponding SM limits at the one-loop level
  are not very different.

\item If low energy SUSY is realised in Nature, then the requirement of
  perturbative gauge coupling unification pushes the inverse radius of
  compactification all the way up to $\sim 10^{10}$ GeV. Thus if
  superpartners of the SM particles are observed at the LHC, the
  nearest KK states within the UED framework are predicted to lie
  beyond the boundary of any observational relevance.
\end{enumerate}

It should be admitted that even if TeV scale extra dimensional theories are
established, the spectrum might be more complicated than what UED
predicts. The confusion is expected to clear up at least when the low-lying KK
states face appointment with destiny within the first few years of the LHC
run. Our intention in the present article has been to choose a simple
framework to study power law evolution.  Flat extra dimensional models are
particularly handy as they provide equispaced KK states which allow an elegant
handling of internal KK summation in the loops. UED is an ideal
test-bed to conduct this study as it has been motivated from various angles
and subjected to different phenomenological tests.

\vskip 5pt

\noindent {\bf{Acknowledgements:}}~ 
We thank E.~Dudas for useful correspondences.  G.~B. thanks CFTP,
Instituto Superior T\'ecnico (Lisbon) and ICTP (Trieste) for
hospitality at different stages of the work.

\end{document}